\documentclass[%
reprint,
superscriptaddress,
showpacs,%
amsmath,amssymb,
aps,
pre,
longbibliography,
]{revtex4-1}

\usepackage[utf8]{inputenc}
\usepackage[T1]{fontenc}
\usepackage[english]{babel}
\usepackage{amsfonts}
\usepackage{graphicx}
\usepackage{color}
\usepackage{hyperref}

\hypersetup{colorlinks = true, allcolors = blue}

\usepackage{mathtools}
\usepackage{amssymb}
\usepackage{xcolor}
\usepackage{xfrac}
\usepackage{esint}

\newcommand{\onefig}[1]{\centering{
    \includegraphics[width=0.99\columnwidth]{#1}}}

\newcommand{\mydd}{\mathrm{d}}
\newcommand{\myee}{\mathrm{e}}
\newcommand{\mat}[1]{\mbox{\boldmath{$\mathrm{#1}$}}}
\newcommand{\const}{\mathrm{const}}
\newcommand{\diagmat}{\mathop{\mathrm{diag}}}

\newcommand{\transp}{\text{T}}
\newcommand{\submin}{\text{min}}

\newcommand{\fbas}{F}
\newcommand{\xbas}{X}
\newcommand{\ybas}{Y}

\newcommand{\xvar}{x}
\newcommand{\yvar}{y}
\newcommand{\uvar}{u}
\newcommand{\vvar}{v}
\newcommand{\jacob}{\mat J}
\newcommand{\xjacob}{\mat J_x}
\newcommand{\ujacob}{\mat J_u}

\begin{document}

\title{Numerical test for hyperbolicity of chaotic dynamics
  in time-delay systems}

\author{Pavel V. Kuptsov}
\email[Electronic address:]{p.kuptsov@rambler.ru}%

\affiliation{Institute of electronics and mechanical engineering, Yuri
  Gagarin State Technical University of Saratov, Politekhnicheskaya
  77, Saratov 410054, Russia}%

\author{Sergey P. Kuznetsov}%

\affiliation{Institute of Mathematics, Information Technologies and Physics, Udmurt
State University, Universitetskaya 1, Izhevsk 426034, Russia}

\affiliation{Kotel’nikov’s Institute of Radio-Engineering and
  Electronics of RAS, Saratov Branch, Zelenaya 38, Saratov 410019,
  Russia}%

\pacs{05.45.-a, 05.45.Jn, 02.30.Ks}

\keywords{hyperbolic chaos, Lyapunov analysis, angle criterion,
  numerical test of hyperbolicity}

\date{\today}

\begin{abstract}
  We develop a numerical test of hyperbolicity of chaotic dynamics in
  time-delay systems. The test is based on the angle criterion and
  includes computation of angle distributions between expanding,
  contracting and neutral manifolds of trajectories on the
  attractor. Three examples are tested. For two of them previously
  predicted hyperbolicity is confirmed. The third one provides an
  example of a time-delay system with nonhyperbolic chaos.
\end{abstract}

\maketitle

Hyperbolic chaotic attractors, like the Smale-Williams solenoid and some
other mathematical examples, manifest deterministic chaos justified in
rigorous mathematical
sense~\cite{Smale67-r01,Anosov-r02,Katok-r03}. In such attractors all
orbits in state space are of saddle type, and their expanding and
contracting manifolds do not have tangencies but can only intersect
transversally. These attractors demonstrate strong stochastic
properties and allow a detailed mathematical analysis. They are
rough, i.e., structurally stable. This means robustness with respect to
variation of functions and parameters in the dynamical equations, and
insensitivity of chaos characteristics to noises, interferences etc.
In the theory of oscillations, since the classic work of Andronov and his
school~\cite{AndrPontr,AndrKhaikVitt}, rough or structurally stable
systems are regarded as those subjected to priority research, and as
the most important for practice. It seems natural that the same should
be true for systems with structurally stable uniformly hyperbolic
attractors. However, until very recently, no physical examples
were known. The hyperbolic attractors were commonly regarded as
purified abstract mathematical images of chaos rather than something
intrinsic to real world systems. In this situation a good way was to
turn to a purposeful construction of systems with hyperbolic dynamics
using a toolbox of physics (oscillators, particles, fields,
interactions, feedback circuits, etc.) instead of that of mathematics
(geometric, algebraic, topological constructions). In this regard,
certain progress has been achieved recently, and many realizable
physically motivated systems with hyperbolic attractors have been
offered~\cite{HyperBook-r04,KuzUFN-r05}.

Naturally, physical and technical devices we deal with are not well
suited to allow mathematical proofs although confident confirmation
of hyperbolicity is significant to exploit properly the relevant
results of mathematical theory. So it is vital to employ
numerical instruments for computational tests of hyperbolicity.

In this respect one has to mention first the so-called cone criterion
based on a mathematical theorem adapted to computer verification that
has been applied for some low-dimensional
systems~\cite{KuzSat-r06,KuzKuzSat-r07,Wilcz-r08}.

The second approach is based on the verification of transversality of
stable and unstable manifolds for orbits belonging to the attractor,
or more generally, to an invariant set of interest. It involves an
inspection of statistical distributions of the angles between the
manifolds. If hyperbolicity holds, all the observed angles have to
be distant from zero. This method was suggested initially in
Ref.~\cite{LaiGeb-r09} and then developed and used with modifications
in~\cite{Hirata-r10,Anishch-r11,Kuznetsov-r12,KuzSelez-r13,
  GinCLV-r14,WolfCLV,KupKuz-r15,FastHyp-r16,CLV2012-r17,Kuznetsov15-r18,
  KuzKrug-r19}. It may be regarded as an extended version of Lyapunov
analyses, well-established and applied successively not only for
low-dimensional systems but for spatiotemporal systems too.

An important class among nonlinear systems with complex dynamics is
formed by systems containing time-delay feedback loops. Such examples
are wide-spread in electronics, laser physics, acoustics and other
fields~\cite{VihLaf-r20}. An adequate mathematical description for
these objects is based on differential equations with
delays~\cite{BellCook-r21,Myshkis-r22,ElsNor-r23}. These dynamical
systems have to be treated as possessing infinite-dimensional state
space since a continuum of data, i.e, a trajectory segment, determines
each new infinitesimal time step. A number of time-delay systems with
chaotic dynamics was
explored~\cite{DorGram-r24,Chrost-r25,Lepri-r26,Mackey-r27,
  Farmer-r28,GrassProc-r29,KuzPonom-r30,Autonom-r31,
  KuzPik-r32,Baranov-r33,KuzKuz-r34}. Several examples of them were
suggested as realizable devices for generation of rough hyperbolic
chaos~\cite{KuzPonom-r30,Autonom-r31,
  KuzPik-r32,Baranov-r33,KuzKuz-r34}. However, no computer
verification of hyperbolicity was provided for these systems as no appropriate
methods were elaborated for this special class of
infinite-dimensional dynamics.

In the present paper we extend the angle criterion to make it
applicable for time-delay systems with one constant retarding time.

To start with, let us recall the content of the angle criterion for
finite-dimensional systems following
Refs.~\cite{FastHyp-r16,CLV2012-r17}. For a system defined by a set
of $m$ ordinary differential equations $\dot\xbas=\fbas(t,\xbas)$ the
corresponding variational equation for infinitesimal perturbations
near the reference orbit $\xbas(t)$ reads
\begin{equation}
  \label{eq:linear_common}
  \dot\xvar=\jacob(t)\xvar,
\end{equation}
where $\xbas,\xvar\in \mathbb{R}^m$ are the state vector and the
perturbation vector, respectively, and
$\jacob(t)\in \mathbb{R}^{m\times m}$ is the Jacobi matrix. The
notation $\jacob(t)$ presumes the dependence of this matrix both on
$t$ and $\xbas(t)$. If the dependence of $\fbas$ and, consequently, of
$\jacob$ on $t$ is explicit, the system is nonautonomous. (In this
case we always restrict ourselves to considering equations with
periodic dependencies of coefficients on $t$.)

Assume that we deal with an invariant set (attractor) possessing $p$
positive, one zero, and $m-p-1$ negative Lyapunov exponents. To
compute the angles between expanding, neutral, and contracting tangent
subspaces we first need to run the standard algorithm for computing
Lyapunov exponents~\cite{Benettin-r35,Shimada79-r36}. The main system
is solved simultaneously with $p+1$ copies of the variational
equation~\eqref{eq:linear_common}. Periodically perturbation vectors
$\xvar$ produced by the variational equations are orthogonalized and
normalized. In matrix notation it corresponds to the so-called QR
decomposition~\cite{GolubLoan-r37} of a $m\times (p+1)$ matrix whose
columns are the perturbation vectors $\xvar$. This procedure
represents the matrix as a product of an orthogonal matrix $\mat Q$
and an upper triangle matrix $\mat R$. Often this is implemented via
the Gram-Schmidt algorithm~\cite{GolubLoan-r37}. The columns of
$\mat Q$ are used in further computations as new perturbation
vectors. After excluding transient iterations, they converge to the
so-called backward Lyapunov vectors~\cite{CLV2012-r17} (they are
called ``backward'' since arrive at $t$ after initialization in the
far past; notice also that the convergence does not imply the time
constancy of these vectors).  Computing Lyapunov exponents, one
accumulates logarithms of the diagonal elements of $\mat R$, but in
the framework of the angle criterion one has to store the backward
Lyapunov vectors instead. The time interval between successive QR
decompositions can be chosen arbitrarily, but it must be not too large
to avoid an overflow of computer digital registers. The backward
Lyapunov vectors are stored for a discrete set of points where the
angles will be computed later.

The next stage of the routine consists in the passage along the same
reference orbit backward in time applying the similar Lyapunov
algorithm but for $p+1$ vectors generated now by the adjoint
variational equation
\begin{equation}
  \label{eq:adjoint_common}
  \dot\yvar=-\jacob^*(t)\yvar.
\end{equation}
Here $\jacob^*(t)$ is the adjoint matrix for $\jacob(t)$, such that
the inner products involving arbitrary vectors $a$ and $b$ satisfy the
identity $\langle \jacob^* a,b\rangle\equiv \langle a,\mat J
b\rangle$. If the inner product is defined as $\langle
a,b\rangle=b^\transp a$, then we have simply
$\jacob^*=\jacob^\transp$, where ``$\transp$'' stands for
transposition. The orthogonal matrices obtained from the QR procedure
in the course of the computations with the adjoint
equation~\eqref{eq:adjoint_common} in backward time converge to the
so-called forward Lyapunov vectors~\cite{CLV2012-r17}.

Now we use the stored forward and backward Lyapunov vectors relating
to the identical trajectory points. Information about the angles is
encoded in a $(p+1)\times (p+1)$ matrix of their pairwise inner
products. The smallest singular value $\sigma_k$ of its top left
$k\times k$ submatrix is the tangency indicator: $\theta_{1}
=\left(\pi/2-\arccos \sigma_{p} \right)$ is the angle between the
expanding subspace and the sum of the neutral and contracting ones,
and $\theta_{2} =\left(\pi/2-\arccos \sigma_{p+1} \right)$ is the
angle between the expanding plus neutral subspace and the contracting
one~\cite{FastHyp-r16}.

One can easily check that for arbitrary solutions to
Eqs.~\eqref{eq:linear_common} and \eqref{eq:adjoint_common} the inner
product remains constant in time, i.e., satisfies the identity
\begin{equation}
  \label{eq:adj_cond}
  \frac{\mydd}{\mydd t} \left\langle \xvar(t),\yvar(t)\right\rangle
  \equiv 0.
\end{equation}
In particular, in the case of nonzero inner product, if one of the
solutions is characterized by a Lyapunov exponent so that
$\xvar(t)\sim \myee^{\lambda t}$, then the other one has a Lyapunov
exponent with the opposite sign, $\yvar(t)\sim \myee^{-\lambda
  t}$. Hence the Lyapunov spectra corresponding to
equations~\eqref{eq:linear_common} and~\eqref{eq:adjoint_common} are
identical up to the signs.

Given only one of Eqs.~\eqref{eq:linear_common}
and~\eqref{eq:adjoint_common}, one can use~\eqref{eq:adj_cond} to find
the other one. This idea can be employed to recover a generic form of
the adjoint variational equation for time delay systems.

Consider a time-delay system
\begin{equation}
  \label{eq:main_eq}
  \dot\xbas=\fbas(t,\xbas,\xbas_\tau),
\end{equation}
where $\xbas\equiv \xbas(t)\in \mathbb{R}^m$,
$\xbas_\tau\equiv \xbas(t-\tau)$, and $\tau$ is the delay time.  The
corresponding variational equation reads
\begin{equation}
  \label{eq:orig_sys}
  \dot\xvar=\xjacob(t)\xvar(t)+\ujacob(t)\xvar(t-\tau),
\end{equation}
where $\xjacob(t)$ and $\ujacob(t)$ are the derivative matrices
composed of partial derivatives of $\fbas$ over components of $\xbas$
and $\xbas_\tau$, respectively.

As a preliminary step to guess a proper form for the adjoint
variational equation we rewrite Eq.~\eqref{eq:orig_sys} as an equation
for a system containing explicitly a delay line, where a signal
propagation is characterized by the time $\tau$:
\begin{gather}
  \label{eq:orig_sys_with_delay_expl}
  \dot\xvar=\xjacob(t)\xvar(t)+\ujacob(t)\uvar(t,L),\\
  \label{eq:delay_expl_for_orig_sys}
  \uvar_t+\uvar_\xi=\delta(\xi)\xvar(t),\\
  \label{eq:delay_expl_sol}
  \uvar(t,\xi<0)=0, \; \uvar(t,\xi>0)=\xvar(t-\xi).
\end{gather}
Here $\uvar\equiv\uvar(t,\xi)$ is the delay line variable, $L=\tau$ is
the length of the delay line, and the subscripts $t$ and $\xi$ stand for
corresponding partial derivatives. The solution to
Eq.~\eqref{eq:delay_expl_for_orig_sys} is a wave
\eqref{eq:delay_expl_sol} propagating in the positive direction from a
source at $\xi=0$. Substituting it
into~\eqref{eq:orig_sys_with_delay_expl} yields the original
equation~\eqref{eq:orig_sys}.

In view of this recasting the variational equation, it is natural
to define the inner product of two perturbation vectors
$\bar{\xvar}=(\xvar,\uvar)$ and $\bar{\yvar}=(\yvar,\vvar)$ as
\begin{equation}
  \label{eq:delay_innver_def}
  \langle\bar{\xvar},\bar{\yvar}\rangle=
  \yvar^\transp(t)\xvar(t)+
  \int\limits_0^L\vvar^\transp(t,\xi)\uvar(t,\xi) \,\mydd\xi.
\end{equation}

The arbitrary solution $\bar{\yvar}$ of desired adjoint variational
equations and $\bar{\xvar}$ satisfying
Eqs.~\eqref{eq:orig_sys_with_delay_expl}-\eqref{eq:delay_expl_sol}
have to fulfill a condition analogous to~\eqref{eq:adj_cond}, i.e.,
$\mydd\left\langle \bar{\xvar},\bar{\yvar}\right\rangle /\mydd t\equiv
0$, with respect to the inner
product~\eqref{eq:delay_innver_def}. This requirement results in the
following form of the adjoint variational equations:
\begin{gather}
  \label{eq:adj_sys_with_delay_expl}
  \dot\yvar=-\xjacob^\transp(t)\yvar(t)-\vvar(t,0),\\
  \label{eq:delay_expl_for_adj_sys}
  \vvar_t+\vvar_\xi=\delta(\xi-L)\ujacob^\transp(t)\yvar(t),\\
  \label{eq:delay_expl_adj_sol}
  \begin{gathered}
    \vvar(t,\xi>L)=0,\\
    \vvar(t,\xi<L)=\ujacob^\transp(t-\xi+L)\yvar(t-\xi+L).
  \end{gathered}
\end{gather}
(One can check directly that the identity
$\mydd\left\langle \bar{\xvar},\bar{\yvar}\right\rangle /\mydd t\equiv
0$ is fulfilled taking into account the equality
$(\mydd/\mydd t) \int_0^Lf(t-\xi)\mydd \xi=f(t)-f(t-L)$.)
Observe that,
in contrast to Eq.~\eqref{eq:delay_expl_for_orig_sys} containing a
source, we introduce a kind of sink
in~\eqref{eq:delay_expl_for_adj_sys}; in other words, we exploit here
an advanced wave solution of
equation~\eqref{eq:delay_expl_for_adj_sys} instead of the usual
retarding wave solution.

Substituting Eq.~\eqref{eq:delay_expl_adj_sol} into
\eqref{eq:adj_sys_with_delay_expl}, one can reformulate the adjoint
variational problem as a differential equation with deviating
argument:
\begin{equation}
  \label{eq:main_adj}
  \dot\yvar=-\xjacob^\transp(t)\yvar(t)-\ujacob^\transp(t+\tau)\yvar(t+\tau).
\end{equation}

The theory of differential equations with deviating arguments
distinguishes three main cases: equations with a retarded argument
(like Eqs.~\eqref{eq:main_eq} and~\eqref{eq:orig_sys}), equations of
neutral type (we do not deal with them here), and equations of leading
or advanced type (like Eq.~\eqref{eq:main_adj})
\cite{BellCook-r21,Myshkis-r22,ElsNor-r23}.  The latter are regarded
as poorly defined with respect to the existence of solutions to
initial value problems. In the context of our study, however, we will
solve such equations in backward time only, so that they behave in a
good way like the equations of retarded type in forward time.

Notice that computing the inner products~\eqref{eq:delay_innver_def} of
vectors obtained as trajectory segments of time duration $\tau$ produced by
Eqs.~\eqref{eq:orig_sys} and~\eqref{eq:main_adj}, one needs to reverse
the order of the adjoint trajectory variables: for
Eq.~\eqref{eq:orig_sys} we set
$[\uvar(t,0),\uvar(t,L)]=[\xvar(t),\xvar(t-\tau)]$, but for
\eqref{eq:main_adj}
$[\vvar(t,0),\vvar(t,L)]=[\yvar(t+\tau),\yvar(t)]$.

We will solve both Eq.~\eqref{eq:main_eq} and the variational
equation~\eqref{eq:orig_sys} numerically with Heun's method that
belongs to the class of second-order Runge-Kutta methods with constant
time step~\cite{NumericalDDE-r38}. The full state vector of the
discrete time approximation of Eq.~\eqref{eq:orig_sys} includes $k+1$
elements $(\xvar_{n-k}, \xvar_{n-k+1},\ldots,\xvar_{n})$, where we
assume $k=\tau/h$ to be an integer, and $h$ is a time step. Numerical
integration of this linear equation implies successive multiplications
of the state vector by a numerical Jacobi matrix $\jacob$ composed of
$[m(k+1)]\times [m(k+1)]$ elements. Regardless of $k$, only a small
number of the matrix elements are nontrivial; most of them are merely
zeros and ones. In particular, for the Heun method this matrix has
$3m^2$ nontrivial elements. In the course of the forward time stage of
the angle computation algorithm one can store these matrix elements
without the risk of exhausting a computer memory. Then, the backward
time stage can be implemented without explicitly solving
Eq.~\eqref{eq:main_adj}. One just computes adjoint matrices from the
stored matrix elements and performs the iterations with them.

Computed in this way, the perturbation vectors converge as $h\to 0$ to
solutions of Eqs.~\eqref{eq:orig_sys} and~\eqref{eq:main_adj} if
the adjoint numerical Jacobi matrix is defined as
\begin{equation}
  \mat \jacob^*=\mat H^{-2}\mat \jacob^\transp \mat H^2,
\end{equation}
where $\mat H=\diagmat(1,\sqrt{h},\ldots ,\sqrt{h})$ and $\mat H^2$ is
a metric tensor generated by a discrete version of the inner
product~\eqref{eq:delay_innver_def};
\begin{equation}
  \label{eq:delay_innver_def_dis}
  \langle\bar{\xvar},\bar{\yvar}\rangle=
  \yvar^\transp_n\xvar_n+
  h\sum\limits_{i=1}^k \vvar^\transp_{i}\uvar_{i}=
  \yvar^\transp \mat H^2\xvar.
\end{equation}
Here $\uvar_i=\xvar_{n-i}$, and if $\vvar_i$ corresponds to an adjoint
vector, the elements of the backward in time solution $\yvar_i$ have to be
taken in the reverse order: $\vvar_i=\yvar_{n+k+1-i}$.

Orthonormalization routines employed in the angle computations have to
be re-implemented according to the nonstandard form of the inner
product~\eqref{eq:delay_innver_def_dis}. When exploiting linear
algebra libraries, a simpler alternative is to change the basis for the
perturbation vectors setting $a'=\mat H a$ with respective
transformation of the Jacobi matrix
\begin{equation}
  \label{eq:modified_jac}
  \mat J'=\mat H \, \mat J\, \mat H^{-1}.
\end{equation}
One can see that
$a^\transp \,\mat H^2 b=(a')^\transp b'$, and
$b=\mat J a$ is equivalent to $b'=\mat J' a'$. In
other words, the transformed Jacobi matrix $\mat J'$ and the corresponding
vectors are written in orthonormal basis whose metric tensor is the
identity. Altogether, computing the Jacobi matrices along a forward
time trajectory we first transform them according to
Eq.~\eqref{eq:modified_jac} and then perform the angle computation
algorithm implementing standard orthonormalization routines without
the need to redefine the inner product.

Now we are ready to apply the angle criterion to particular time-delay
systems.

The first is a nonautonomous system based on the van der Pol
oscillator of natural frequency $\omega_0$ supplied with a
specially designed time-delay feedback~\cite{KuzPonom-r30}:
\begin{equation}
  \label{eq:kuzponom}
  \ddot{\xbas}-[A\cos(2\pi t/T)-\xbas^2]\dot{\xbas}+\omega_0^2\xbas=
  \epsilon \xbas_\tau\dot \xbas_\tau\cos\omega_0t.
\end{equation}
The parameter controlling the oscillator excitation is modulated with
period $T$ and amplitude $A$. Accordingly, the oscillator alternately
manifests activation and damping. If the retarding time $\tau$ is
properly tuned, say $\tau =\frac{3}{4} T$, the emergence of
self-oscillations at each next stage of activity is stimulated by a
signal emitted at the previous activity stage. Since the delayed
signal is squared and mixed with auxiliary oscillations of frequency
$\omega_0$, the stimulating force has again frequency $\omega_0$,
but the doubled phase in comparison with the original oscillations. As a
result, we get a sequence of oscillation trains with phases at
successive excitation stages obeying the doubly-expanding circle map
that is a chaotic Bernoulli-type map
\begin{equation}
  \label{eq:bernoulli}
  \phi_{n+1}=2\phi_{n}+\const \mod 2\pi.
\end{equation}
According to argumentation in Ref.~\cite{KuzPonom-r30}, this means that
the attractor for a Poincar\'{e} map, which corresponds to states
obtained stroboscopically at $t_n =nT$, is of Smale-Williams type,
and the respective chaotic dynamics is hyperbolic. (In fact, the
system~\eqref{eq:kuzponom} has an additional attracting fixed point
$\xbas= 0$, but its basin of attraction is very narrow. Arbitrarily
chosen initial conditions of relatively large amplitude provide an
approach of orbits to the Smale-Williams attractor.)

\begin{figure}
  \onefig{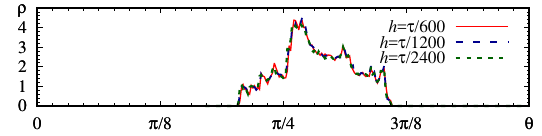}
  \caption{\label{fig:angdist_kuzponom}(color online) Distributions of
    $\theta_1$ for system~\eqref{eq:kuzponom} at different time steps
    $h$, see the legend. $T=6$, $\tau=\frac{3}{4}T$, $A=4.7$,
    $\epsilon=0.3$, $\omega_0=2\pi$. The first four Lyapunov exponents are
    $0.114$,$-0.139$,$-0.714$, $-0.801$.}
\end{figure}

\begin{figure}
  \onefig{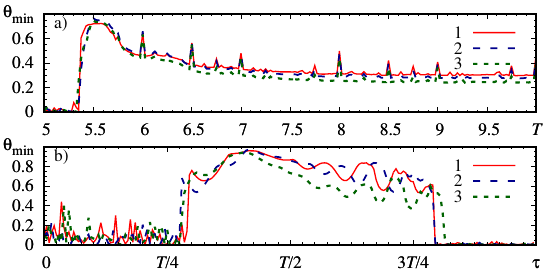}
  \caption{\label{fig:angmin_kuzponom}(color online) The minimum cutoff
    angle $\theta_\submin$ against (a) $T$ and (b) $\tau$. (a)
    $\tau=3T/4$. Curve 1: $A=4.7$, $\epsilon=0.3$; 2: $A=5$,
    $\epsilon=0.25$; 3: $A=5.5$, $\epsilon=0.2$. (b) Curve 1: $A=4.7$,
    $\epsilon=0.3$, $T=6$; 2: $A=5$, $\epsilon=0.2$, $T=6.5$; 3:
    $A=5.2$, $\epsilon=0.15$, $T=7.5$. For both panels
    $\omega_0=2\pi$.}
\end{figure}

The system~\eqref{eq:kuzponom} has a single positive Lyapunov exponent and
the others are negative. (Zero Lyapunov exponent is absent because of the
nonautonomous nature of the system.) Thus, we need to implement the
angle computation routine with $p=1$.

Figure~\ref{fig:angdist_kuzponom} shows histograms of the angle
distributions computed for numerical solutions of the equations with
different time steps $h$. The angles are obtained for the
corresponding stroboscopic map at $t=t_0+nT$, where $t_0$ is large
enough to exclude transients, and $n=0,1,2,\ldots$. Regardless of $h$
(being sufficiently small) the distributions have a well reproducible
form that supports the correctness of our definitions in the continual
limit. The form of the histograms certainly confirms the hyperbolicity
of chaotic dynamics in the system. Indeed, the distribution is well
separated from zero angles.

Figure~\ref{fig:angmin_kuzponom} shows the dependencies of the minimum
cut-off angle $\theta_\submin$ against the system parameters to verify
robustness of the hyperbolic chaos. In its panels (a) and (b) $T$ and
$\tau$ are varied, respectively, for three arbitrarily chosen sets of
parameters. Observe that the angle distributions remain well separated
from the origin in a wide parameter domain. (The spikes on the curves
in panel (a), e.g., at $T=6,6.5,7$ etc., correspond to situations when
the angle distributions are maximally distant from zero, although it
is not easy to explain the origin of the phenomenon clearly.) In the
left part of the plot (a) relating to the case $\tau=\frac{3}{4} T$
one can see a narrow domain where the minimal angles are close to
zero, and the dynamics becomes nonhyperbolic. In panel (b) an
extensive domain of hyperbolicity in the middle part shows that the
fine tuning of $\tau$ at given $T$ is actually not necessary because
the hyperbolicity persists in a relatively wide range.

The second system we consider is an autonomous model with an attractor of
Smale-Williams type suggested in Ref.~\cite{Autonom-r31}, see
also~\cite{HyperBook-r04,Arzh-r39}:
\begin{equation}
  \label{eq:autonom}
  \begin{aligned}
    \dot \xbas&=-\omega_0
    \ybas+\tfrac{1}{2}\mu(1-\xbas^2_\tau-\ybas^2_\tau)
    \xbas+
    \epsilon \xbas_\tau \ybas_\tau,\\
    \dot \ybas&=\phantom{-}\omega_0\xbas+\tfrac{1}{2}
    \mu(1-\xbas^2_\tau-\ybas^2_\tau)\ybas.
  \end{aligned}
\end{equation}
Here $\mu$ is an excitation parameter and $\omega_0$ can be treated as
a natural frequency. When $\epsilon=0$, this system generates trains of
oscillations of frequency $\omega_0$ periodically alternating with
damping stages of very low amplitude. Due to the term providing a
delayed feedback controlled by the small parameter $\epsilon$, each new
train of oscillations arises from a seed signal produced by the system
at the previous excitation stage. Because of the quadratic
nonlinearity of the respective term, each time the phase is doubled in
comparison with the previous one. As a result, the phases at successive
excitation stages evolve according to the Bernoulli-type
map~\eqref{eq:bernoulli}, and, as argued in Ref.~\cite{Autonom-r31},
this means that the Poincar\'{e} map constructed for the
system~\eqref{eq:autonom} has an attractor of Smale-Williams
type. Following~\cite{Autonom-r31}, we will consider the Poincar\'{e}
map on the surface $\xbas^2+\ybas^2=1$ taking into account the
passages of phase trajectories in the direction of increasing
amplitude and computing the angles between subspaces there.

\begin{figure}
  \onefig{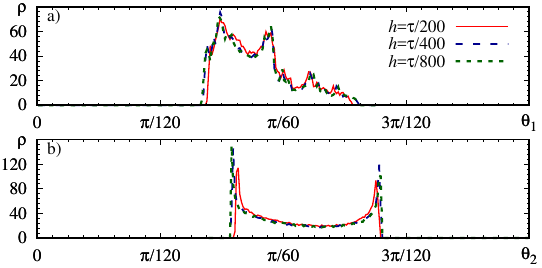}
  \caption{\label{fig:angdist_autonom}(color online) Distributions of
    the first (a) and second (b) angles for the Poincar\'{e} map of
    the autonomous system~\eqref{eq:autonom} computed at different time
    steps, see the legend. $\mu=1.6$, $\omega_0=2\pi$,
    $\epsilon=0.05$, $\tau=2$. The first four Lyapunov exponents are
    $0.060$, $0$, $-1.938$, $-2.463$.}
\end{figure}

\begin{figure}
  \onefig{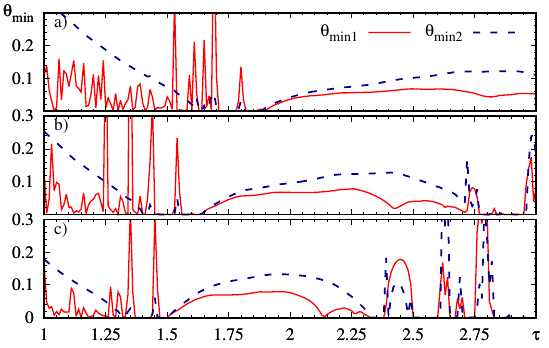}
  \caption{\label{fig:angmin_autonom}(color online) First and second
    minimum cutoff angles vs. delay time $\tau$ at various
    parameters: (a) $\mu=1.6$, $\epsilon=0.05$; (b) $\mu=1.8$,
    $\epsilon=0.07$; (c) $\mu=2$, $\epsilon=0.03$. For all panels
    $\omega_0=2\pi$.}
\end{figure}

Autonomous system~\eqref{eq:autonom} has a single positive Lyapunov
exponent, a zero one, and the other exponents are negative. The zero
Lyapunov exponent and the corresponding neutral subspace vanish for
the Poincar\'{e} map, so that one can consider only the angles between
the expanding and contracting subspaces as above. However,
implementing this approach, one needs to project the computed
perturbation vectors onto the section surface and then evaluate the
angles between these projections. A simpler way is to consider
perturbation vectors as they appear in computations in full phase
space and to check both the angles $\theta_1$ (the expanding subspace
vs. the sum of neutral and contracting subspaces) and $\theta_2$ (the
expanding plus neutral subspace vs. the contracting subspace). Chaos
is hyperbolic when both of these angles never vanish.

Figure~\ref{fig:angdist_autonom} shows histograms of the distributions
of these angles obtained for certain parameters of the system (see the
caption) at different values of the integration step $h$. We see a
good convergence as $h\to 0$ both for $\theta_1$ and for $\theta_2$
(panels (a) and (b), respectively). Both distributions are well
separated from the zero angle having clearly expressed cutoffs that
confirms hyperbolicity of the attractor. The robustness of this regime is
illustrated in Fig.~\ref{fig:angmin_autonom}(a-c), which represents the
dependence of the minimum cutoff angles $\theta_{\submin1}$ and
$\theta_{\submin2}$ versus the parameter of time delay $\tau$ for
three sets of other parameters. On all three plots well-defined ranges
occur corresponding to domains of existence of the hyperbolic
attractor where both angles are well detached from zero.

The last example we present is aimed to outline the difference between the
hyperbolic and nonhyperbolic chaos. It is the well-known Mackey-Glass
system~\cite{Mackey-r27,Farmer-r28,GrassProc-r29}
\begin{equation}
  \label{eq:mckglass}
  \dot \xbas=a \xbas_\tau/\left(1+\xbas_\tau^{10}\right)-b\xbas,
\end{equation}
where $a=0.2$ and $b=0.1$. When $\tau>17$, chaotic oscillations occur
in the system~\cite{Farmer-r28,GrassProc-r29}.

To perform an adequate comparison with the previous cases, we have to
introduce for this system an appropriate Poincar\'{e} section
providing well-expressed separation of the successive crossings
by phase trajectories.
A good and satisfactory expedient is based on
complementing the model~\eqref{eq:mckglass} with an auxiliary equation
$\tau\dot \ybas=\xbas-\xbas_\tau-\ybas$ and locating the section
surface at $\ybas=0$. The variable $\ybas$ roughly follows $\xbas$ but
smoothes out its high-frequency fluctuations.

\begin{figure}
  \onefig{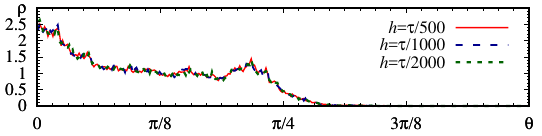}
  \caption{\label{fig:angdist_mckglass}(color online) Distribution of
    $\theta_1$ for system~\eqref{eq:mckglass} at different time steps,
    see the legend. $\tau=23$. The first four Lyapunov exponents are
    $0.0096$, $0$, $-0.0114$, $-0.0350$.}
\end{figure}

Figure~\ref{fig:angdist_mckglass} shows histograms for the angle
$\theta_1$ between the expanding subspace and the sum of neutral and
contracting subspaces for the attractor observed at $\tau=23$. The
attractor has one positive, one zero, and other negative Lyapunov
exponents. Observe the clearly expressed violation of hyperbolicity:
the distributions demonstrate a significant probability of angles close
to zero, which implies occurrence of tangencies for the subspaces. It is
a sufficient condition to judge about the absence of
hyperbolicity, so that there is no need to analyze a distribution for
$\theta_2$.

To conclude, we have developed the method of hyperbolicity
verification based on the angle criterion for time-delay systems,
which may be regarded as an extension of numerical Lyapunov
analysis. Three particular examples have been tested. For two of them
the previously believed hyperbolicity has been confirmed and for the
third one the nonhyperbolic nature of the generated chaos has been
established. We have restricted ourselves here to considering systems with one delay
time. Extension for two or more delays requires further elaboration of
the algorithm.

As regards the theoretical foundation and general formulation of the
method, the work was supported by RSF grant No 15-12-20035 (SPK). The work of
elaborating computer routines and numerical computations
for particular examples was supported by RFBR grant
No~16-02-00135 (PVK).

\newpage
\bibliography{hypdelay}

\end{document}